# Method for Creating and Detecting Hydrogen Sorption Sites Using Gamma Radiation

*Barbara G. Muga and M. Luis Muga*[1]

TOFTEC INC, PO Box 15358, Gainesville, FL 32604

## Abstract

Using gamma radiation and volumetric analysis of desorbed gas, hydrogen gas bonding sites have been created and detected in select materials. Desorption of hydrogen was followed over a benign temperature-pressure range. The extent of active site formation depends on radiation dosage; quenching of sites occurs over prolonged heating at low pressures. An estimate of the hydrogen bonding energy can be made on the basis of a partial temperature profile of the gas released at one atmosphere pressure. It appears that the bonding energy can be adjusted by mixing candidate materials. A guide for further investigation and application of the method is outlined.

## Introduction

Hydrogen is attractive as an energy reservoir because of its high specific energy content and the scarcity of environmental pollutants generated on combustion in air. Unfortunately, one serious obstacle impedes its general replacement of conventional fossil fuel sources, *viz*., the storage of hydrogen in a safe, economical, and easily retrievable compact form. Approaches to overcoming this barrier involve research in numerous directions, among which are 1) high pressure storage, 2) cryogenic storage as liquid, 3) formation of metal and mixed metal hydrides and 4) adsorption and absorption on various materials that have been treated or modified to enlarge the active interior or surface area. After over a decade of a widely publicized effort, these approaches have not led to practical implementation of the desired goal. The thrust of this work is to introduce an alternate method of forming and detecting hydrogen sorption sites, useful over an easily attainable temperature-pressure range, as a prelude to further investigation.

Gamma irradiation creates non-equilibrium conditions in solids *via* the Compton effect[1]. The net result is that the electronic environment at surface and interior sites is activated by this penetrating radiation. Gamma irradiation creates a variety of active sites some of which may attract hydrogen molecules with strength between physisorption and true (or partial) chemical bonding (or chemisorption). This intermediate strength attraction is necessary to effect a capability to absorb and release hydrogen over a practical range of temperature and pressure values. The ideal attractive strength has been suggested[2] as having a heat of absorption of *ca*. -15 kJ/mol. For comparison, physisorption, or van der Waals attraction, involves about 6 kJ/mol $H_2$, true chemical bonds, such as exist in hydrocarbons, have a bond strength of over 400 kJ/mol, and the typical hydrogen bond, as exists in liquid water, reflects an energy of about 10 kJ/mol. This method may be used to explore the effect of gamma radiation on mixtures of select materials in a combinatorial-like approach and as a guide for determining the characteristics of the bonding sites thus formed as discussed in the final section.

[1] Corresponding author: muga@chem.ufl.edu

A practical goal[3] for $H_2$ storage is a material with minimum weight percent sorption of about ten percent and can be defined by the formula:

$$Mass\% = \left[ mR / \left( mR + M \right) \right] * 100 \qquad (1)$$

where m is the molecular mass (2 amu) of sorbed $H_2$ gas, M is the molecular mass (in amu) of the host group and R is the number ratio of $H_2$/host sites. For example, if the host group has a mass of 90 amu, R must be 5/1 in order to sorb at ten weight percent. It is clear that a combination of small M and large R is favored. Therefore we initially focus our attention on material of low atomic mass composition. By varying the mixture and adding known impurities, the effect on $H_2$ sorption for a given radiation dose will serve as a guide to the elemental composition needed for the generation of $H_2$ host sites. Thereafter, determination of the structural basis of the host group must be made by other means.

## Experimental

The experimental method used is direct and quite simple in principle, though somewhat tedious and time consuming. The results reported herein were done in two phases over a period of about five years as the procedural details were modified to improve accuracy and reproducibility. The experimental conditions were confined to less than 8 bar pressure and a temperature range from 273 K to 473K in order that any observed sorbing sites would potentially be useful for practical applications. A diagram of the experimental setup is shown in Figure 1.

A sealed cartridge filled with 3-5 grams of a select mixture of powders is attached to a manifold equipped for evacuation and $H_2$ or He gas pressurization. Macro sized samples are preferred over microgram quantities that might lead to misinterpretation due to sorbed impurities. Separate valving allows for controlled depressurization as the contained gasses are vented into an inverted graduated burette and collected over water at 1 bar pressure. After initial material pre-treatment (including extensive ball milling) and cartridge evacuation, a series of cycles is run. In a first phase of experiments each cycle consisted of the following steps at room temperature unless otherwise stated: **(a)** cartridge pressurized to 8 bar pressure, **(b)** optional irradiation from a 0.5 MCi Co-60 source for over 1000 hours, **(c)** cooling of cartridge and contents to 0 °C followed by depressurization and collection of exhaust gas at 1 bar pressure, **(d)** heating of cartridge to 100 °C or higher at one bar pressure and collection of exhaust gas. In a second phase of experiments, step **(d)** was modified by heating the cartridge and contents in increments of 10 to 25 °C, and the off-gasses were collected and measured after each incremental rise in temperature. Each experimental cycle covers both the loading (adsorption/absorption) and release (desorption) of $H_2$ (or He) gas under the cited conditions. For each starting situation (with or without radiation exposure) experimental cycles were repeated to assure reproducibility of measurements. Thus an experimental series began with 2 cycles using helium (as a systems check) followed by up to 12 cycles using $H_2$ gas. Duplicate samples were irradiated and processed in each experiment.

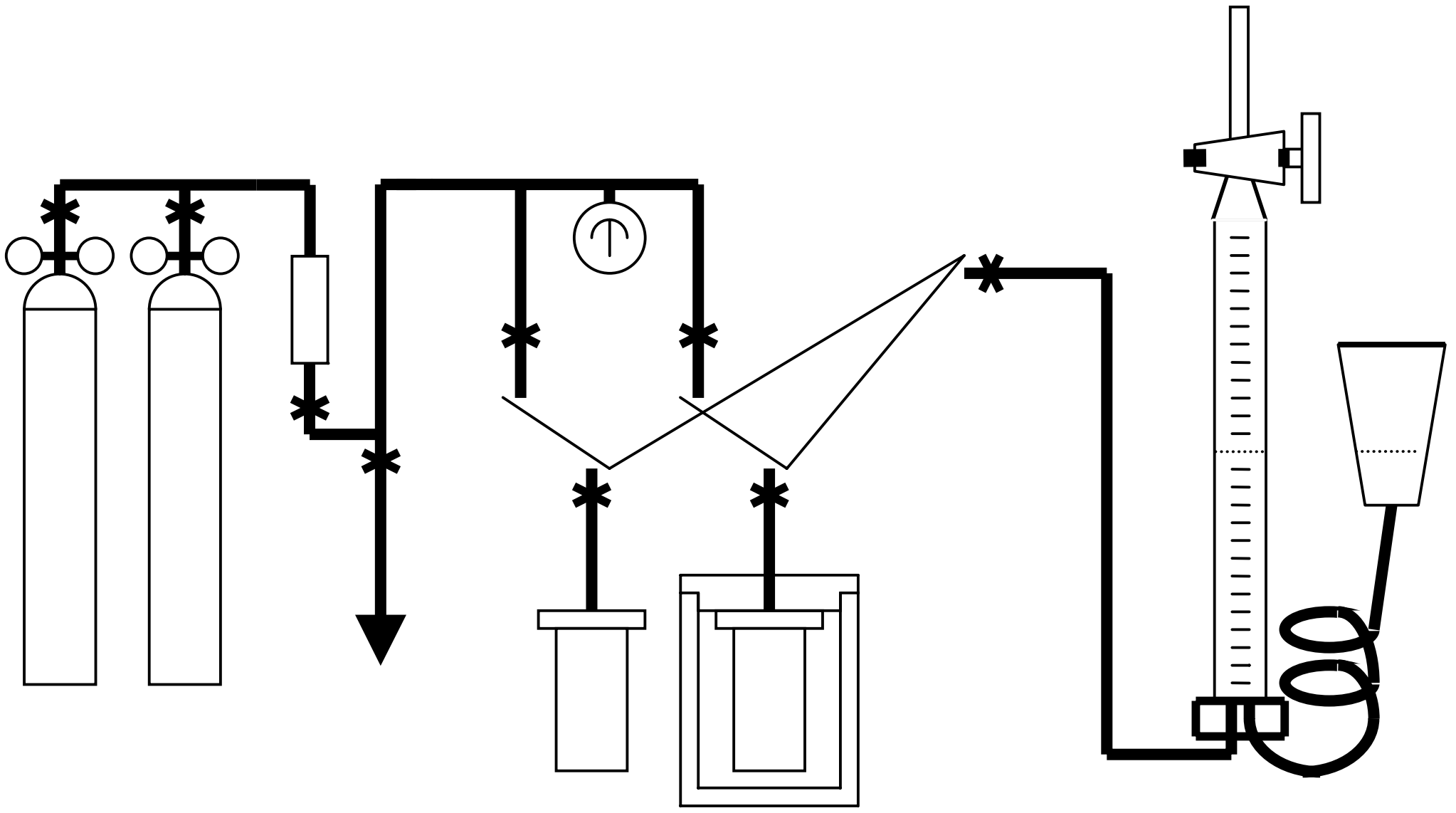

**Figure 1.** Schematic diagram, not to scale, of experimental setup for introducing gas followed by volumetric analysis of desorbed gasses under controlled T, P conditions. On left, pressurized He and $H_2$ tanks connected through drier to manifold (with gauge) and vacuum system (arrow). On right, inverted burette and leveler bulb. Center, detachable sample cartridges with Dewar flask or oven.

Measured gas volumes were corrected for water vapor content and converted to *NTP* (20 °C, 1 bar) volumes. Off-gas volumes measured in this first phase represent the expansion in $H_2$ gas volume at constant 1 bar pressure (due to a 100 °C temperature rise) *plus* any release of sorbed $H_2$ gas. The free-space-volume in the cartridge may be calculated from the T, P conditions of depressurization. The off-gas *volume increase* due solely to temperature related expansion of the free-space-volume gas can be calculated; it is better determined experimentally on a given sample by running identical experimental cycles *before* irradiation, 2 cycles with He and 2 cycles with $H_2$ gas. Subtracting this volume expansion (temperature related) of gas in the free-space-volume from the measured volume increase (after irradiation) results in the volume of gas released from true (gamma-created) sorption sites over the 0 – 100 °C incremental rise in temperature at 1 bar pressure. This component of off-gas, *if observed*, confirms that gamma irradiation creates effective $H_2$ gas sorption sites. Although mass spectrometric identification of the recovered gas was not made, the gas was assumed to be $H_2$ for the following reasons: **(a)** on loading in successive cycles, only $H_2$ gas was introduced and cycles were reproducible, **(b)** the recovered gas was insoluble in the burette water and **(c)** no change of pH was observed as recovered gas bubbled up into the inverted burette. Cycles repeated after initial irradiation also serve to establish differential effects such as quenching of sites over time, and as a result of heating. Partial temperature profiles of sorbed gas release were recorded from 273 to 373 K and higher in order to estimate the associated enthalpy of desorption. Complete desorption at still higher temperatures was avoided in order to reduce the possibility of host site quenching that would adversely affect subsequent cycles.

## Results and Discussion

In most samples heretofore run, no physisorption or chemisorption was indicated whatsoever, however in a few select materials a significant and measurable amount of $H_2$ gas was sorbed and released. Figure 2 summarizes the results of the initial phase of this work. These results indicate that **(a)** practical $H_2$ gas bonding and desorption host sites are possible and can be generated by gamma irradiation of select materials, **(b)** at least two different active sites may be formed, one of which anneals rapidly (in just about an hour) with heating, the second one of which is stable at 100 °C, **(c)** active site formation depends on the extent of irradiation dose, **(d)** active site formation depends on make-up of the irradiated mixture and hence on elemental, molecular and structural composition, **(e)** sorption and desorption of $H_2$ gas can be effected under relatively benign conditions of temperature and pressure changes and can be repeated, and **(f)** prolonged heating under vacuum ($10^{-6}$ torr) serves to quench the active sites.

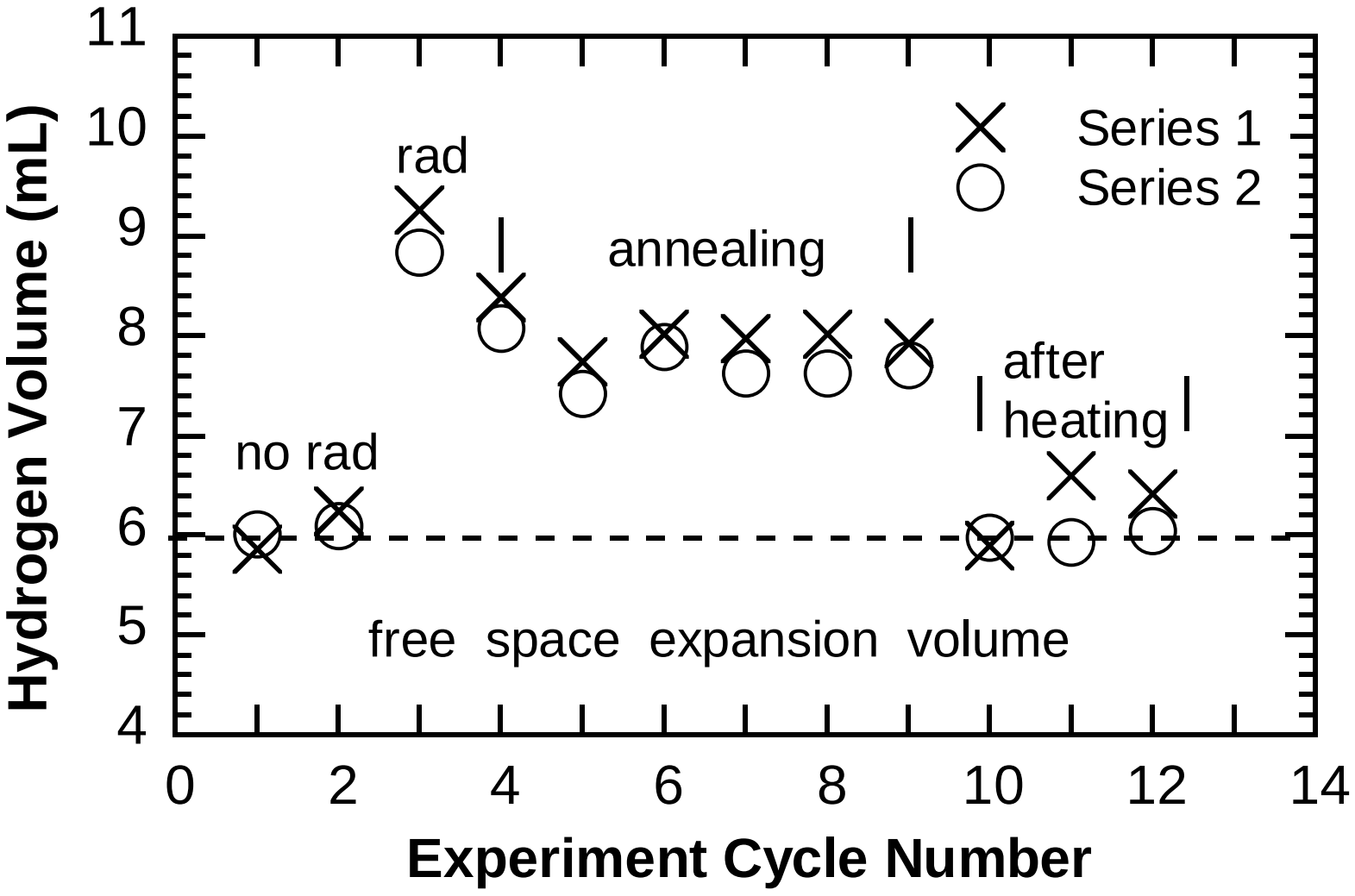

**Figure 2.** $H_2$ recovered per each loading/release cycle as explained in text. Cycles #1 and #2 – free space expansion volume recovery at 100 °C and 1 bar pressure ***before gamma irradiation***. Cycle #3 – $H_2$ recovery ***after gamma irradiation*** of 1300 hrs. Volume recovery above dashed line represents $H_2$ recovery from sorbing sites. Volume below dashed line represents free space expansion volume of non-sorbed $H_2$ gas. Cycles #4 -#9 – $H_2$ recovery after annealing at 1 bar pressure $H_2$ for ~ 1 hour at 100 °C at end of previous cycle and reloading. Cycles #10 -#12 – $H_2$ recovery after prolonged heating of 6 hours at 100 °C under vacuum of $10^{-6}$ torr suggesting almost total quenching of sorbing sites has occurred.

Unusual bonding sites are expected in gamma irradiated material because of the locally intense energy deposited by Compton electron passage and ionization. While electronic excitation has little effect on the gross mechanical properties of most materials (e.g. strength, elasticity, compressibility), lattice atoms, stripped of their electrons, may form active centers that are attractive to one or more invasive species. It is this possibility that is exploited. The Co-60 irradiator available for this work[4] is capable of delivering a dose rate (initially) of about 0.5 Mrad/hour. Thus a fifty-day exposure will result in a total dosage of about 600 Mrad. [One rad is the absorbed dose that deposits 100 ergs per gram, or 6.24 x $10^{13}$ eV/g sample.] Converting this dose to energy deposition in a sample (say, with an average atomic mass of 12 amu) results in a value of about 0.75 eV for every potential host atom. For reference, the ionization energy for a hydrogen atom is 13.6 eV. This energy deposition is not uniformly distributed and much of it dissipates harmlessly. Most host atoms will receive little if any energy yet a sufficient number (a guesstimate of 0.1 % at this dosage) should be candidates for active site formation. For a three gram sample of low atomic mass material one might expect to recover about 0.25 mmol of $H_2$ gas, if each active site adsorbs one hydrogen molecule. This is "in the ball park" figure of the amount that we are currently recovering (~0.13 mmol, or ~3 mL $H_2$, see Figure 2) from *select materials*. Ergo, we see that gamma radiation is indeed an effective tool for creating unusual bonding host sites in certain materials from which recovery of gas absorbate can be made *over a benign temperature/pressure range*.

In a second phase, partial temperature profiles at constant 1 bar pressure provided the basis for estimating the desorption enthalpies. For example, the select material was irradiated for 4330 hours after establishing pre-irradiation calibrations. Measured volumes were corrected for water vapor content and converted to normal temperature-pressure (*NTP*) conditions of 293 K and 1 bar. Figure 3 displays a profile before and after irradiation; the experimental points were fitted with a $4^{th}$ degree polynomial and plotted as continuous lines for further analysis. The *difference* between these two curves represents the volume V of absorbate recovered from gamma-created sorption sites, *i.e.* the desorption curve, and is shown as Figure 4a. Working with the fitted curve, the *differential* of this desorption curve, dV/dT, is also shown (Figure 4b) as temperature is increased.

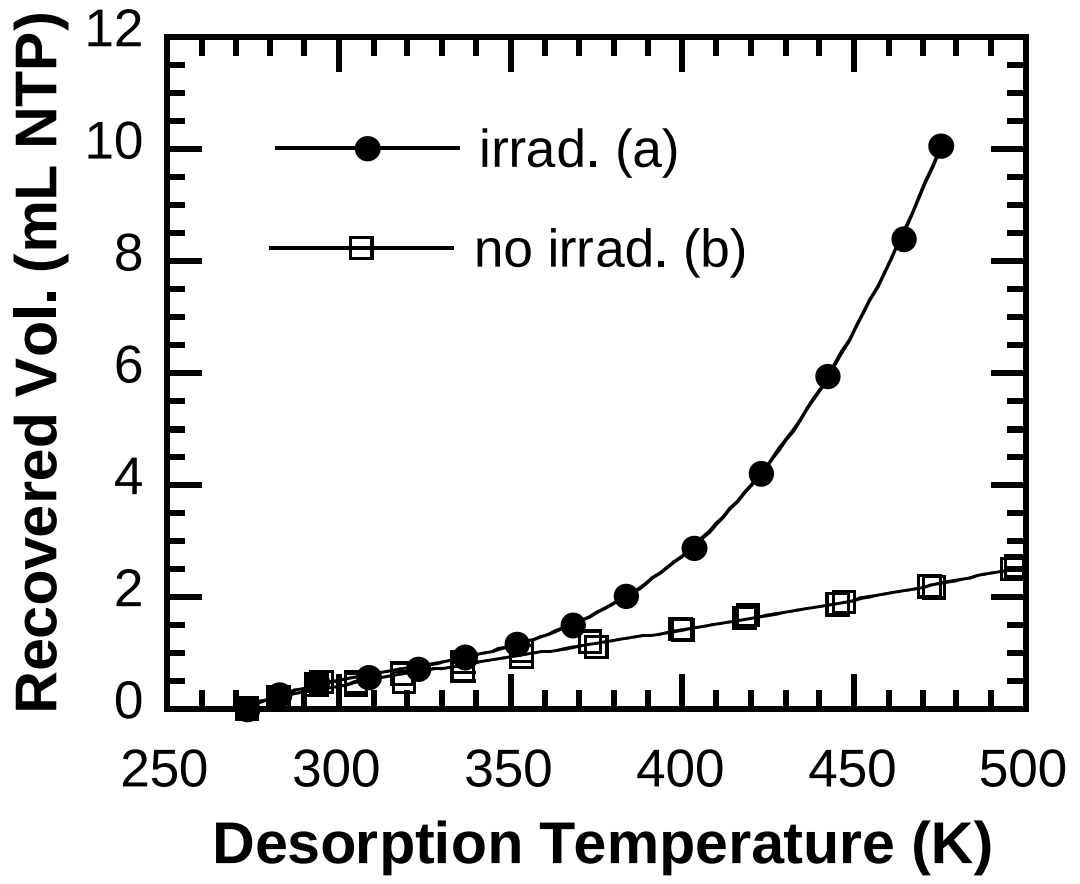

**Figure 3**. Volume of recovered $H_2$ as temperature is increased, (a) after irradiation and (b) before irradiation. Solid lines represent $4^{th}$ degree polynomials fitted to experimental points.

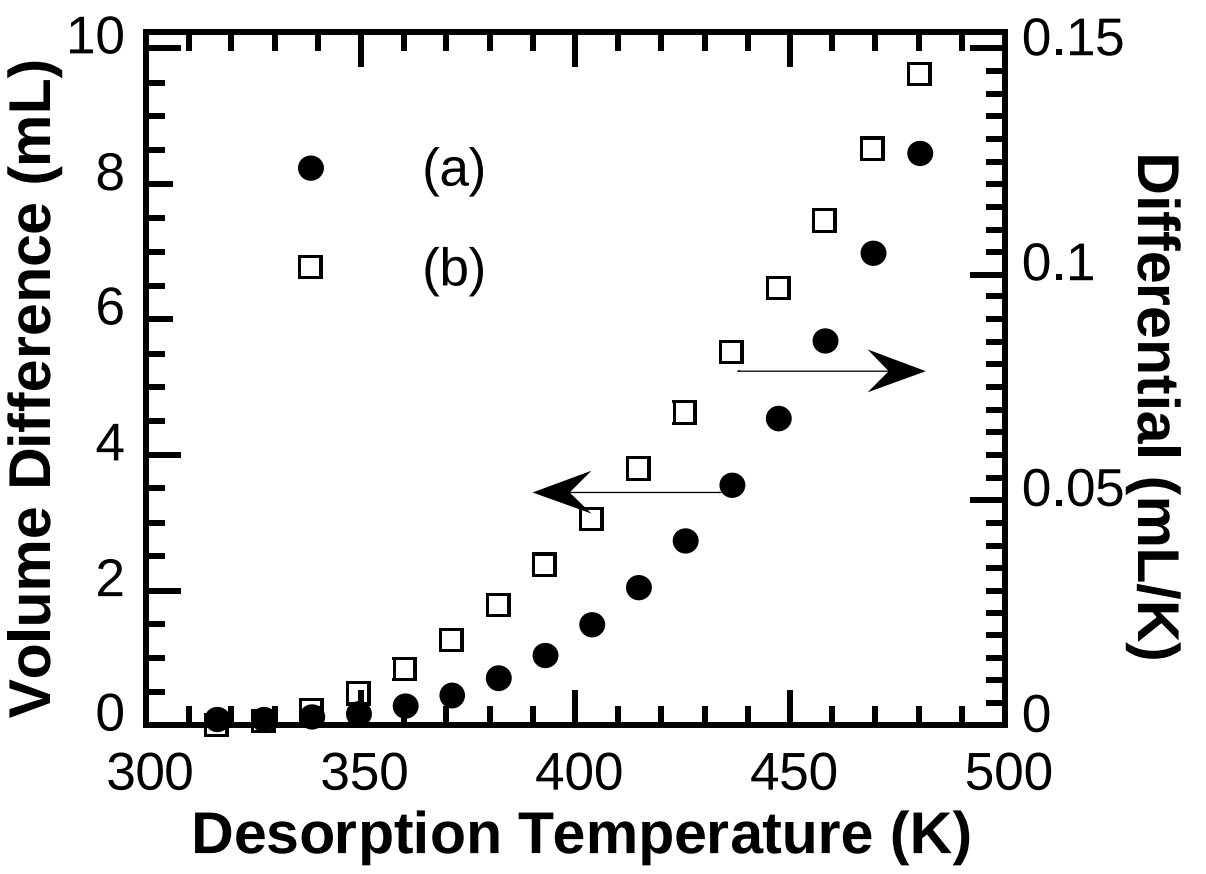

**Figure 4**. (a) Volume of desorbed $H_2$ and (b) differential or rate of volume recovery of desorbed $H_2$, as function of temperature after irradiation.

At the highest temperature (473 K) it is evident that not all absorbate has been released as the desorption curve has yet to level off to constant value and the differential plot covers less than half of the expected "bell shaped" (or gaussian-like) curve, having not yet reached the apex at the highest recorded temperature. Estimates of the maximum recoverable volume $V_{max}$ (corresponding to the total number of gamma created sites, $N_\gamma$) and the volume not recoverable $\Delta V$ (due to initially vacant sites) can still be made and serve as a useful guide in completing the determination of desorption enthalpy as described next.

Following the approach of Langmuir[5] for mono-layer chemisorption, let $N_\gamma$ be the number of sorption sites of which $N_s$ are occupied by absorbate molecules and $N_v$ are vacant at temperature T and constant pressure P of 1 bar. Then $N_\gamma = N_s + N_v$. Consider the equilibrium system:

$$S \cdot H_2(solid) + \Delta H_d \underset{k_s}{\overset{k_d}{\rightleftarrows}} S(solid) + H_2(gas) \tag{2}$$

where S represents the substrate, $\Delta H_d$ is the desorption enthalpy (negative of adsorption/ absorption enthalpy), and $k_d$ and $k_s$ are respectively, the temperature sensitive desorption and sorption rate constants. At *constant* 1 bar pressure and at a given temperature, the forward (desorption) rate $R_d$ is expected to be proportional to the number of occupied sites $N_s$ and the reverse (sorption) rate $R_s$ is proportional to the number of vacant sites $N_v$. Hence at equilibrium the forward and reverse rates are equal:

$$R_d = k_d N_s = R_s = k_s N_v \tag{3}$$

and the equilibrium constant $K_{eq}$ (at 1 bar pressure) is given by:

$$K_{eq} = \frac{k_d}{k_s} = \frac{N_v}{N_s} \tag{4}$$

But the number of sites $N_i$ from which absorbate is recovered corresponds to a *NTP* volume $V_i$ as approximated by the ideal gas law:

$$N_i = N_A P_{ntp} V_i / RT_{ntp} = cV_i \tag{5}$$

where $N_A$ is the Avogadro number, R is the gas constant (8.317 J/mol K) and, $P_{ntp}$ and $T_{ntp}$ are 1 bar and 293 K, respectively, and c becomes the appropriate constant of proportionality.

If the beginning temperature (of profile) is too high, the recovered volume V at any higher temperature may be short by $\Delta V$, the volume corresponding to the number of initially vacant sites at the starting value, 273 K, of temperature profile. In the case of Figure 4b, this deficit appears to be negligible as suggested by the gradual departure from baseline for the differential curve; an abrupt beginning would suggest that not all sorbing sites are occupied at this point. Hence, we make the association:

$$N_v = c\left(V + \Delta V\right) \tag{6}$$

Likewise, the number of occupied sites $N_s$ will be equal to $N_\gamma$ (proportional to the maximum recoverable volume $V_{max}$ from completely saturated substrate with $N_\gamma$ sites) *less* the number of vacant sites $N_v$ (proportional to the *already* recovered volume from an initially saturated substrate) giving the relation:

$$N_s = N_\gamma - N_v = c\left[V_{max} - \left(V + \Delta V\right)\right] \tag{7}$$

Substituting these relations into the van't Hoff isochor[6] we have:

$$\ln K_{eq} = \ln\left[\frac{V+\Delta V}{\left[V_{\max} - \left(V+\Delta V\right)\right]}\right] = -\frac{\Delta H_d}{R}\left(\frac{1}{T}\right) + \text{constant} \tag{8}$$

where $V_{max}$ and $\Delta V$ are adjustable constants consistent with values implied by Figure 4. Plotted as ln $K_{eq}$ vs. (1/T), a straight line is expected with slope $-\Delta H_d/R$ after adjusting the constants $V_{max}$ and $\Delta V$ to give a best fit.

The select sample of Figures 3 and 4 resulted in the plot shown in Figure 5a. The validity of $\Delta H_d$ is dependent on the proper choice of $V_{max}$ and $\Delta V$. Since desorption was not complete, the value of $V_{max}$ must be estimated from the beginning shape of the differential curve and the last recorded (highest T) point of the difference curve (Figures 4a,b). For example, the differential plot of Figure 4b falls well short of reaching the apex of its expected bell-like shape, suggesting that the recovered amount (8 mL) at highest T represents only 1/4 to 1/3 of the maximum recoverable volume $V_{max}$. Hence, the first trial choice for $V_{max}$ is roughly 3 to 4 times the recovered volume (plus $\Delta V$) at final temperature. Best fit adjustable constants $V_{max}$ and $\Delta V$ were found to be about 26 mL and 0.0 mL respectively, values that are consistent with expectations. From the slope of the line a value of 44 kJ/mol was obtained for $\Delta H_d$, a value somewhat higher than the ideal value of 15 kJ/mol. On reloading this sample and running another temperature profile, a $\Delta H_d$ value of about 69 kJ/mol resulted, but the recovery cycle was marred by an unexplained low yield of recovered gas. A third cycle on this sample resulted in a value of 38 kJ/mol. Unfortunately, a second identically treated sample was compromised during processing.

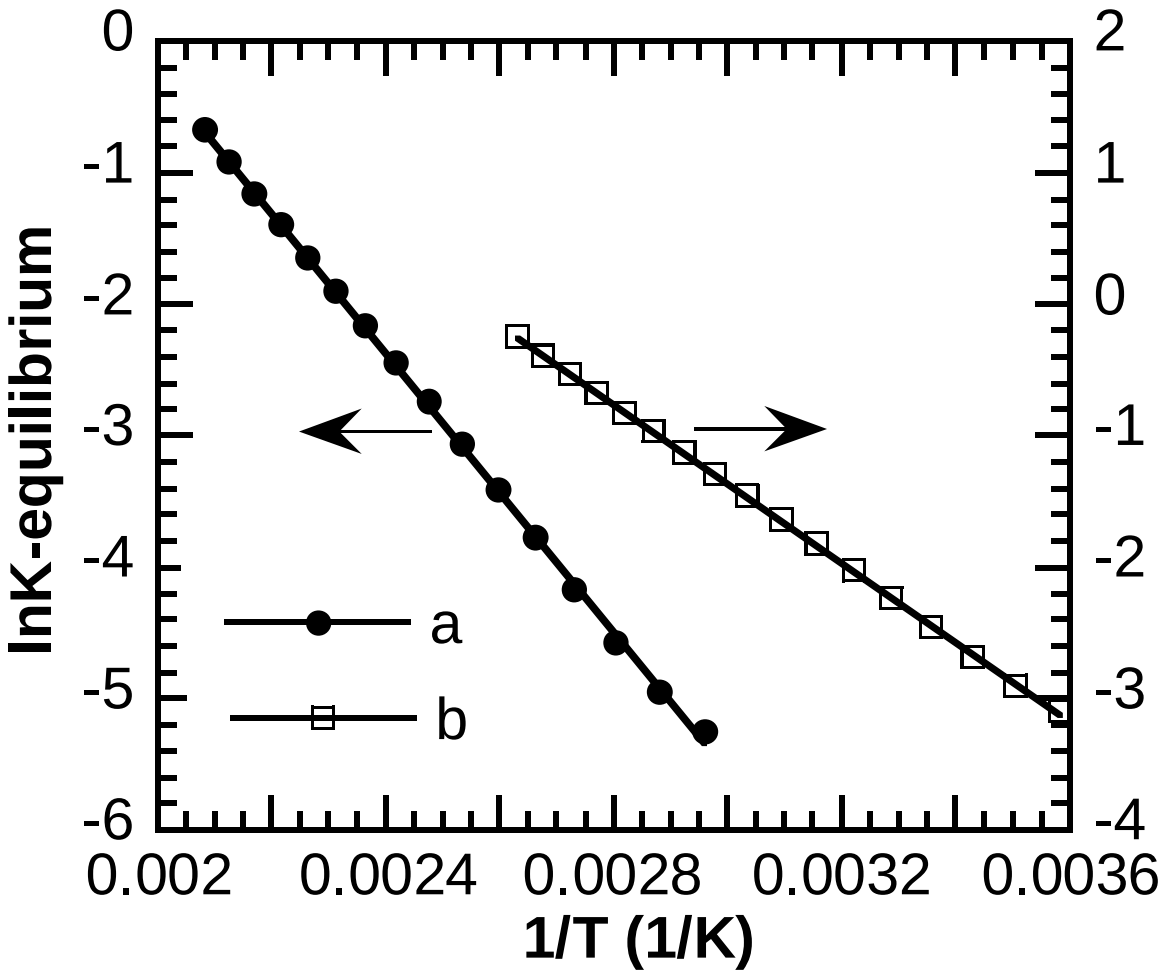

**Figure 5.** van't Hoff plot showing straight line dependence (with arbitrary y-axis offset) for (a) sample A and (b) mixture A+B.

In a companion experiment, the select material was mixed with another candidate material and, after calibration steps, exposed to gamma irradiation for 1510 hours. A representative temperature profile set is shown in Figure 6. Difference and differential plots are shown in Figure 7 for the first cycle following irradiation. Notice the contrast between the differential plots shown in Figures 4b and 7b. The abrupt beginning edge of the differential plot of Figure 7b suggests that host sites are not all occupied at the lowest recorded temperature (273 K). Here the choice of $V_{max}$ is more clear as the differential (expected bell-shaped) curve nears but does not quite reach the apex; a good first choice being about 2-3 times the 2.6 mL recovered volume (plus smaller $\Delta V$) at T equal 373 K. Best fit adjustable constants for the van't Hoff plot, shown as Figure 5b, were $V_{max}$, 6.5 mL, and $\Delta V$, 0.22 mL. The $\Delta H_d$ value for the first cycle (after irradiation) was determined to be 25 kJ/mol and the value determined for the second cycle was 24 kJ/mol. A second identical sample resulted in $\Delta H_d$ values of 24 kJ/mol and 23 kJ/mol respectively for the first and second cycles after irradiation.

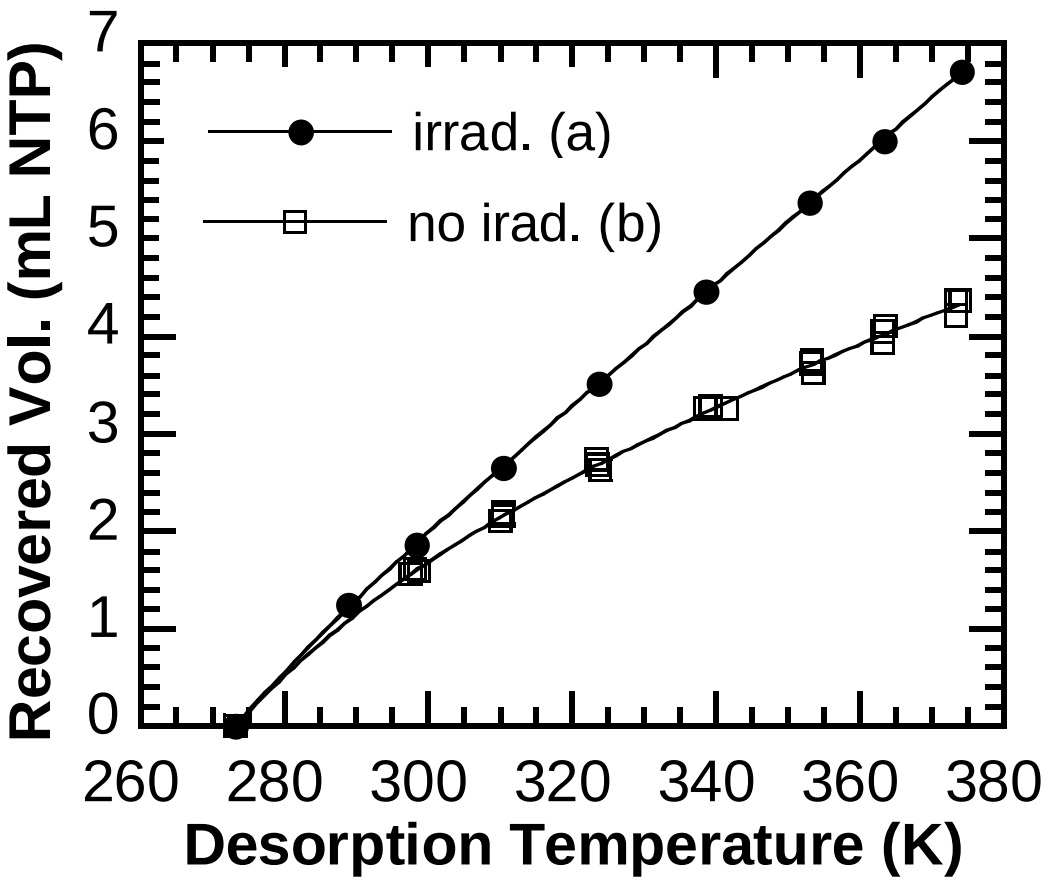

**Figure 6**. Volume of recovered $H_2$ from mixed sample (a) after irradiation and (b) before irradiation, as temperature is increased. Solid lines represent $4^{th}$ degree polynomials fitted to experimental points.

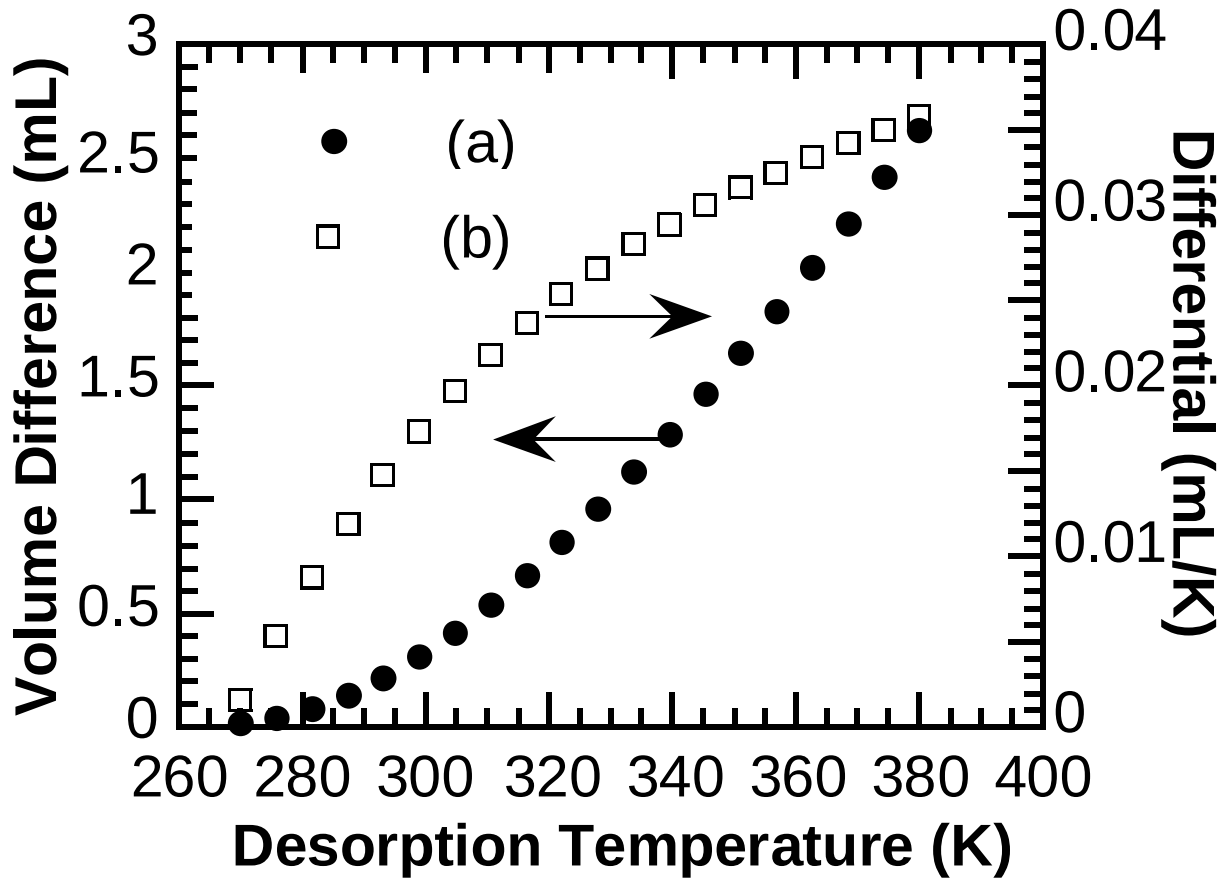

**Figure 7.** (a) Volume of desorbed $H_2$ from mixed sample and (b) differential or rate of recovery of desorbed $H_2$ from mixed sample, as temperature is increased after irradiation.

Table 1 lists the enthalpy estimates for all samples. The listed ranges in $\Delta H_d$ values reflect only the variation about the weighted best fit value for changes in $V_{max}$ or $\Delta V$ of $\pm$ 25%. Since the calculated enthalpy values are based on a portion only (1/4 to 1/2) of a complete desorption profile, the values of $\Delta H_d$ may have a wider range of uncertainty, perhaps $\pm$10% overall. Comparing these values we may conclude that the strength of $H_2$ attraction depends on the material mix and may be significantly lowered or tailored to choice. Moreover, it is possible that two or more host sites of different attractive strengths are present. It would have been of interest to follow up with additional cycles, ending with a final cycle at a high enough temperature so as to effect complete desorption in order to better estimate $V_{max}$, and hence $\Delta H_d$. However, at this point in time lab operations were suspended due to downsizing of laboratory facilities.

TABLE 1: Values of $\Delta H_d$ determined for cycles following $\gamma$-irradiation

| Sample | Cycle #1 | Cycle #2 | Cycle #3 |
|---|---|---|---|
| A | 44±2 kJ/mol | (69) kJ/mol | 38±1 kJ/mol |
| A+B | 25±2 kJ/mol | 24±2 kJ/mol | - |
| A+B | 24±2 kJ/mol | 23±3 kJ/mol | - |

## Conclusions

The thrust of this project was to explore the possibility for loading (sorption) and release (desorption) of measurable amounts of $H_2$ gas under benign T-P conditions using gamma irradiated powders as substrate. A simple volumetric analysis was used to confirm that sorption at 0 °C, 8 bar and desorption over a range 0-100 °C, at 1 bar pressure are practicable limits that were ***experimentally*** achieved for select materials. The enthalpy of desorption has been found to be within striking distance of the reported ideal value and appears to be adjustable by choosing mixtures of select materials. By screening a variety of related mixed samples, the material composition best suited to forming $H_2$ sorption/desorption sites may be found. It is not suggested that gamma radiation be used in a fabrication process to mass produce hydrogen storage materials, rather we are using gamma radiation as a screening tool to form detectable sorption host sites using 3-5 g samples. The host sites have been pre-selected by this screening technique to have the properties that are desirable for practical $H_2$ storage and release. Determining the effectiveness of site formation as composition or mix of the target material is varied can serve as a guide to the ultimate goal of identifying the exact elemental and molecular arrangement that constitutes a useful $H_2$ sorption/desorption host site. Instrumental analysis with ESR, NMR, STM, *etc*. may help to elucidate the nature of the host sites. In this manner, determination of important parameters such as R and M (see eq. 1) will indicate the potential of the material for further development and the optimum elemental ratio can be determined for guiding a program aimed at duplicating the responsible bonding arrangement by ordinary chemical synthesis. Thus identification of the sorbing host site and its elemental composition and structure can be followed by attempts at chemical synthesis of bulk amounts.

Finally, it should be remarked that although this work emphasized $H_2$ as the detecting gas of sorbing host sites, the method may be applied to other gasses. For example, binding strengths of gamma-excited host sites could be determined for He, other GVIII gasses, $N_2$, $CH_4$, CO, *etc*.

## Acknowledgment

This work was supported indirectly and in part by the U.S. Social Security Administration through the courtesy of the more senior author (MLM). We wish to thank Professor Emeritus R. J. Hanrahan and the Chemistry Department, University of Florida, for arranging access to the University of Florida Co-60 irradiator facility.